\documentclass[runningheads]{llncs}
\usepackage{graphicx}

\usepackage{amsmath}
\usepackage{amsfonts}
\usepackage{booktabs} 
\usepackage{textcomp}  %
\usepackage[toc,page]{appendix}

\begin{document}
\title{Smart Meter Data Anomaly Detection using
Variational Recurrent Autoencoders with Attention}
\titlerunning{Smart meter data anomaly detection using VRAE with attention}
\author{Wenjing Dai\inst{1} 
\and
Xiufeng liu\inst{1} 
\and
Alfred Heller\inst{2}
\and
Per Sieverts Nielsen\inst{1}}

\authorrunning{Wenjing et al.}
\institute{Department of Technology, Management and Economics, \\
Technical University of Denmark, 2800 Kgs. Lyngby, Denmark 
\email{\{weda,xiuli,pernn\}@dtu.dk}
\and
Niras, \O stre Havnegade 12, 9000 Aalborg, Denmark \\
\email{ahr@niras.dk}
}

\maketitle
\begin{abstract}
In the digitization of energy systems, sensors and smart meters are increasingly being used to monitor production, operation and demand. Detection of anomalies based on smart meter data is crucial to identify potential risks and unusual events at an early stage, which can serve as a reference for timely initiation of appropriate actions and improving management. However, smart meter data from energy systems often lack labels and contain noise and various patterns without distinctively cyclical. Meanwhile, the vague definition of anomalies in different energy scenarios and highly complex temporal correlations pose a great challenge for anomaly detection. Many traditional unsupervised anomaly detection algorithms such as cluster-based or distance-based models are not robust to noise and not fully exploit the temporal dependency in a time series as well as other dependencies amongst multiple variables (sensors). This paper proposes an unsupervised anomaly detection method based on a Variational Recurrent Autoencoder with attention mechanism. with ``dirty" data from smart meters, our method pre-detects missing values and global anomalies to shrink their contribution while training. This paper makes a quantitative comparison with the VAE-based baseline approach and four other unsupervised learning methods, demonstrating its effectiveness and superiority. This paper further validates the proposed method by a real case study of detecting  the anomalies of water supply temperature from an industrial heating plant.

\keywords{Anomaly detection \and Variational autoencoder \and Smart meter data \and Attention mechanism.}
\end{abstract}

\section{Introduction}
\begin{figure}[htp]
\vspace{-18pt}
    \centering
    \includegraphics[width=0.9\textwidth]{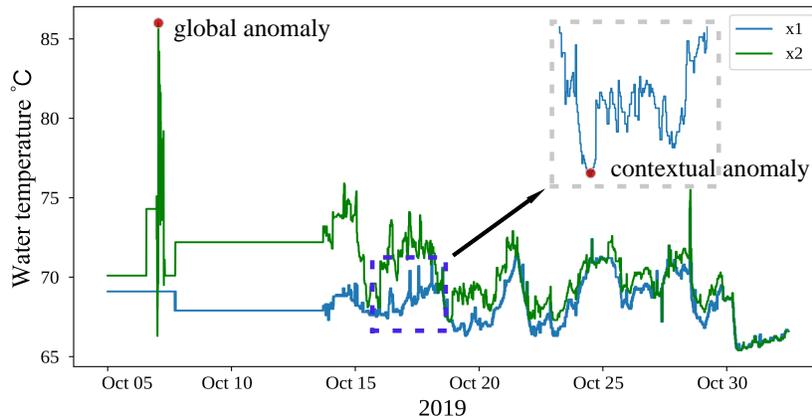}
    \vspace{-15pt}
    \caption{A fragment of the water supply temperature data set in our paper, with global and contextual anomalies marked as red dots.}
    \label{fig:anomalies}
    \vspace{-18pt}
\end{figure}

To achieve sustainable development, effective management of production, distribution, transport and consumption of smart energy systems has become a focus for researchers and engineers \cite{lund2017smart}. As the operations of energy systems can be disrupted by various events such as equipment failures, power outages and malfunctions, energy systems have started to use Internet of Things (IoT) sensors and smart meters for monitoring and automation. Therefore, anomaly detection in smart meter data plays an important role in ensuring the healthy operation of an energy system. When performing anomaly detection, three types of anomalies are widely detected: global, contextual and collective anomalies \cite{2009Anomaly}. In this paper, we mainly focus on global and contextual anomalies (defined in Section~\ref{section:definition}) from smart meter data. Global and contextual anomalies may indicate equipment failures or wrong operations. These two types of anomalies detection are needed for providing early warnings, thus reducing or avoiding economic losses. Figure~\ref{fig:anomalies} illustrates examples of these two types of anomalies in smart meter data. 

However, using smart meter data to detect anomalies faces some key challenges. First, smart meter data are the time series data from production or consumption, and are characterised by different seasonal patterns and highly nonstationary. Various patterns and nonstationary data require more generic and robust anomaly detection methods. Second, smart meter data are typically of high volume, high dimensionality and lack of labeled anomalies, which necessitates the use of unsupervised or semi-supervised approaches. In addition, there are many data quality issues for the collected data, such as missing values, outliers and temporal inconsistency. How to deal with ``dirty" data will affect the performance of results.

Currently, there are several algorithms for detecting anomalies in energy data, such as \cite{chahla2019novel,hollingsworth2018energy,liu2018,pereira2018unsupervised}, but these algorithms are mainly designed to detect point anomalies and do not distinguish between global anomaly and contextual point anomaly. In addition, irregularly missing points should also be considered as anomalies, which occur very frequently in the time series of smart meters due to transmission or meter failures. It is therefore necessary to develop an effective and reliable anomaly detection model for smart energy systems.

In this paper, we propose an unsupervised anomaly detection algorithm for global and contextual anomalies in smart meter data by using variational recurrent autoencoders (VRAE) with attention. This algorithm can work without labels and takes advantage of the pre-detected global anomalies while training. It can also take advantage of the occasional labels when they are available.

\begin{itemize}
    \item We adapt and extend VRAE models by taking into account anomaly detection for smart meter data. The proposed model is capable of detecting not only global anomalies but also contextual anomalies. Although the model is presented for smart meter data, it can also be applied to other time series with time dependency.
    
    \item We propose the method for minimising the impact of global anomalies and missing points on latent variables in the model training, using linear interpolation and an improved evidence lower bound function, which can improve the model performance.
    \item We evaluate the method comprehensively by comparison with other baseline methods using a synthetic data set; and present a real world case study for the proposed method.
\end{itemize}

\section{Related Work}

\subsection{Traditional anomaly detection methods}
Traditional anomaly detection methods include the traditional statistical approaches, e.g., \cite{7836693,jakkula2010outlier,5175339,seem2007using,zhang2011anomaly,6520712}, the clustering-based approaches, e.g., \cite{chen2011energy,zhao2014adaptive}, the prediction-based approaches, e.g., \cite{liu2016online,liu2020,liu2018}, the nearest neighbour approaches, e.g. \cite{breunig2000lof,jakkula2010outlier,kriegel2009loop}, the dimensionality reduction approaches, e.g. \cite{7836693,scholkopf1998nonlinear} and other complementary models. These approaches can show good performance and effectiveness for their specific applications. However, due to the wide variation of energy data such as patterns, domain expert effort is often required to select a suitable detector for a particular type of anomaly. In addition, since most existing methods have their constraints or limitations in terms of parameterisation, interpretability and generalisability, a detection framework based on ensemble learning cannot even help to achieve better results.

\subsection{Unsupervised deep learning models}
A rich body of literature presents unsupervised learning algorithms for detecting anomalies using deep learning techniques, among many others, which include \cite{fan2018analytical,malhotra2016lstm,malhotra2015long,xu2018unsupervised,zhang2019deep}. Deep learning approaches can be further categorised into predictive models \cite{malhotra2015long}, VAE \cite{an2015variational}, Generative Adversarial Networks (GAN) \cite{creswell2018generative} and VRNN \cite{solch2016variational}. 
For modeling sequential data such as time series, Recurrent Neural Networks (RNNs) show their advantage over others because of their capability to model long-term temporal dependence. RNNs (e.g. the Long and Short Term Memory (LSTM) \cite{hochreiter1997long} and the Gated Recurrent Unit (GRU)) introduce the so-called internal self-looping states in the network, which can accumulate information from the past. \cite{santolamazza2018anomaly} combined ARIMA and LSTM to train a prediction model for energy anomaly detection. In this paper, we introduce LSTM into our neural network architecture for modeling the temporal dependence of time series.

VAE has been successfully applied in several applications for anomaly detection tasks, including \cite{pol2019anomaly,su2019robust,xu2018unsupervised,zhang2019time}. %
Hollingsworth et. al. \cite{hollingsworth2018energy} proposed an autoencoder-based ensemble method to detect anomalies in building energy consumption data and evaluated their performance among reconstruction ability, high-level feature quality and computation efficiency. Compared to autoencoders, the variational inference technique \cite{goodfellow2016deep} implements the encoding of the latent space as a distribution and enables the probabilistic reconstruction of a single generated value by a probabilistic model \cite{an2015variational}. However, in the field of smart energy, few applications have previously used generative models to detect anomalies. Existing work based on VAE is not designed for energy smart meter data and requires domain experts in detectiong different types of anomalies. %

\subsection{Attention mechanism for deep learning models}
Attention mechanisms \cite{bahdanau2014neural,luong2015effective,vaswani2017attention} have been introduced to obtain state-of-the-art performance when modeling sequences such as natural language processing. Attention mechanism can model the relationship regarding different positions of a single sequence or across multiple sequences to obtain representative sequences. For example,  Pereira et. al. \cite{pereira2018unsupervised} used weighted sum of all encoder hidden states as the attention, which are then fitted to the decoder. The attention mechanism can, therefore, tackle the weakness of processing a long sequence by neural networks. However, there are still limited attempts and their application in anomaly detection for energy time series data which exists temporal interdependency at different time positions. 

\section{Problem statement and Proposed Method}

\subsection{Problem statement}
\label{section:definition}
Given historical data of n-dimensional time series with length $T$, i.e. $X=\left(\mathbf{x}_{1}, \cdots, \mathbf{x}_{n}\right)^{T} \in \mathbb{R}^{n \times T}$, our method is capable of detecting two types of anomalies:

(a) \textbf{Global anomalies}: given an input time-series $X$, a global anomaly is a timestamp-value pair $\left\langle t, x_{t}\right\rangle$ where the observed value is far from the rest of the data.

(b) \textbf{Contextual anomalies}: given an input time-series $X$, a contextual anomaly is a timestamp-value pair $\left\langle t, x_{t}\right\rangle$ where the observed value differs significantly from its neighbours in the same context, but is not a global anomaly.

\subsection{Proposed method}

\subsubsection{Global anomaly detection and labeling.}
Data collected from real world applications are often dirty, which require preprocessing before being used for analysis. The training process for the anomaly detection should ideally learn from ``normal" data, rather than learn from abnormal data. One of challenges of unsupervised anomaly detection methods is how to minimise the impact of abnormal data as much as possible. Hence, we detect global anomalies and sequential missing points and label them as anomalies before training. We use a statistical method based on histograms of each dimension. For multivariate time series with $n$ dimensions, we first construct a univariate histogram with $k$ bins for each dimension. Second, the frequency of samples in each histogram (dimension) is used as a density estimate of those samples. The higher the score of a sample, the higher the probability of anomaly.

For the missing points, we categorise them into the following two categories:  single missing values and sequential missing values. For single missing values, we fill them with synthetically generated values using linear interpolation. For sequential missing values, the imputation error for missing data is accumulated according to the length of the missing subsequences. As it is difficult to generate sequential data that follow their original patterns, we therefore fill these sequential missing values with zeros and label them as anomalies.

\vspace{-15pt}
\subsubsection{Network architecture and implementation.}
\begin{figure*}[htp]
    \centering
    \includegraphics[width=0.95\textwidth]{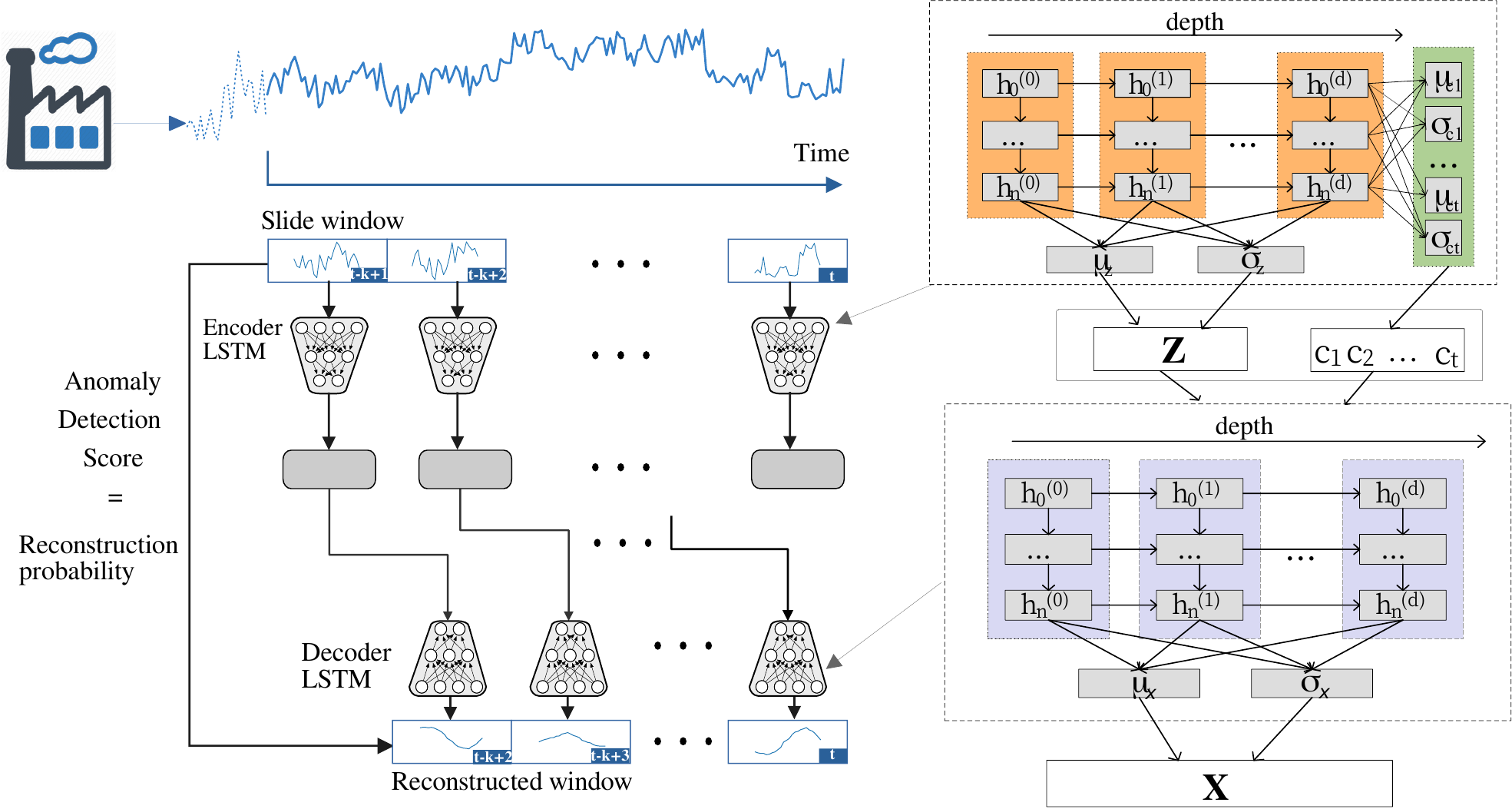}
    \vspace{-12pt}
    \caption{The network architecture of the proposed model.}
    \label{fig:structure}
    \vspace{-20pt}
\end{figure*}

Figure~\ref{fig:structure} shows the overall neural network architecture of the proposed model. As shown in the figure, multivariate time series data come from smart meters of industries.  Given multivariate time series $X$, we first use a sliding window with length $W$ to segment the time series into subsequences e.g. ($x_{t-W+1}, \ldots, x_{t}$). The subsequences are then used as the input of the proposed model which uses a variational auto-encoder architecture with LSTM to learn normal patterns from training data. The right side of Figure~\ref{fig:structure} shows the detailed network structure with attention mechanism. The network structure is a variational recurrent auto-encoder which is composed of an encoder and a decoder. 
\vspace{-18pt}

In the VRAE, the encoder compresses the input time series into the fixed-length latent representation $z$ based on the variational distribution $q_{\phi}(z \mid x)$ and outputs the hidden states $\mathbf{h}_{t}$ as the summary of the past information until the time at $t$. The latent variables $z$ are drawn from a distribution with a given prior $p_\theta (z)$, which is usually a multivariate unit with Gaussian distribution $\mathcal{N}(0, \mathrm{I})$. Here, we assume the prior distribution of the latent variables $z$ as a multivariate normal distribution, $p_\theta (z) \sim \mathcal{N}(0, \mathrm{I})$. The outputs of 
the encoder are the parameters ($\boldsymbol{\mu}_{z}$ and $\boldsymbol{\sigma}_{z}$) for the posterior $q_{\phi}(z \mid x)$. The approximate posterior $q_{\phi}(z \mid x)$ of $z$ is diagonal Gaussian $q_{\phi}(z \mid x)\sim \mathcal{N}\left(\boldsymbol{\mu_{\mathrm{z}}}, \boldsymbol{\sigma_{\mathrm{z}}}^{2} \mathrm{I}\right)$, where the mean $\boldsymbol{\mu_{\mathrm{z}}}$ and the co-variance $\Sigma_{\mathrm{z}}=\boldsymbol{\sigma_{\mathrm{z}}}^{2} \mathrm{I}$ are derived from the two fully connected layers ($\boldsymbol{\mu_{\mathrm{z}}}$ and $\boldsymbol{\sigma_{\mathrm{z}}}$ layers in figure~\ref{fig:structure}) with Linear and SoftPlus activations, respectively. The latent variable $\mathbf{z}$ (chosen to be K dimensions) are then sampled from the approximate distribution with reparameterization trick, $ \mathbf{z}= \boldsymbol{\mu_{\mathbf{z}}} +\boldsymbol{\sigma_{\mathrm{z}}} \cdot \boldsymbol{\epsilon}$, where $\boldsymbol{\epsilon} \sim \operatorname{Normal}(\mathbf{0}, \mathbf{I})$ is an independent random variable used for feasible stochastic gradient descent. The decoder also uses a LSTM network to reconstruct the data from latent variable $\mathbf{z}$ through the generation distribution $p_{\theta}(x \mid z)$, and outputs the parameters ($\boldsymbol{\mu_{\mathbf{x}}}$ and $\boldsymbol{\sigma_{\mathrm{x}}}$) of $p_{\theta}(x \mid z)$.

The objective of a VAE is to maximise the evidence lower bound (ELBO), $\mathcal{L}\left(\theta, \phi ; \mathbf{x}\right)$, which can be written as follows:

\begin{equation}
\begin{aligned}
\label{eqn: elbo}
\log p_{\theta}(\mathbf{x}) & \geq \mathcal{L}\left(\theta, \phi ; \mathbf{x}\right) \\
&=E_{q_{\phi}\left(\mathbf{z} \mid \mathbf{x}\right)}\left[\log p_{\theta}(\mathbf{x} \mid \mathbf{z})\right]-\mathcal{D}_{\mathrm{KL}}\left(q_{\phi}(\mathbf{z} \mid \mathbf{x}) \| p_{\theta}(\mathbf{z})\right)
\end{aligned}
\end{equation}
where the $\phi$ and $\theta$ are the parameters of the encoder and decoder, respectively. The first item of the right-hand side of the equation is the reconstruction loss, which can be approximated by Monte Carlo integration \cite{an2015variational}. The second item  $D_{K L}$ is the Kullback-Leibler (KL) divergence between the approximate posterior and the prior distribution of the latent variable $z$.

To tackle the posterior collapse in the variational inference and the weakness in a long sequence, we additionally apply self-attention mechanism that promotes interaction between the inference model and the generative model. The attention model extracts a context vector based on all hidden states encoded from the input time series. The LSTM encoder computes all hidden states $\left\{\mathbf{s}_{i}\right\}_{i=1}^{T_{x}}$ from the input time series, while the LSTM decoder estimates the hidden state $\mathbf{h_t}$ at each time $t$ by a recurrent function using the previous hidden state $\mathbf{h_{t-1}}$ and the context vector, denoted by:
\vspace{-2pt}
\begin{equation}\mathbf{h}_{t}=f\left(\mathbf{h}_{t-1}, \mathbf{c}_{t}\right) \quad \text { where } \quad \mathbf{c}_{t}=\sum_{i=1}^{T_{x}} \alpha_{t i} \mathbf{s}_{i}\end{equation}
where $\mathbf{c}_{t}$ is the context vector containing the weighted sum of all source hidden states $\mathbf{s}_{i}$ encoded from the input time series. The attention weights, $\boldsymbol{\alpha}_{t}=\left\{\alpha_{t i}\right\}_{i=1}^{T_{x}}$ , are computed by the score function of measuring the similarity between the hidden states $\mathbf{s}_{t}$ at time t in the encoder and all hidden states $\left\{\mathbf{s}_{i}\right\}_{i=1}^{T_{x}}$ of the last recurrent layer in the encoder. The self-attention models the relevance of each pair of the hidden states of different time instances in the encoder. Here, we use the scaled dot-product similarity \cite{vaswani2017attention} as the score function because of its high learning efficiency for a large input.

Due to the bypass phenomenon \cite{bahuleyan2017variational},  the variational latent space may not learn much due to the powerful attention mechanism.  We therefore use the variational attention mechanism to model context vectors as probability distributions. We choose the prior distribution of the context vectors $ \mathbf{c}_{t}$ as the Gaussian standard distribution, i.e., $\mathbf{c}_{t} \sim \operatorname{Normal}(\mathbf{0}, \mathbf{I})$. We do the same for the latent variables. The encoder first computes the deterministic context vector $\mathbf{c}_{t}=\sum_{i=1}^{T_{x}} \alpha_{t i} \mathbf{s}_{i}$, then passes it to the linear layers to compute the parameters of the approximate posterior $q_{\boldsymbol{\phi}}^{(a)}\left(\mathbf{c}_{t} \mid \mathbf{x}\right) \sim \operatorname{Normal}(\boldsymbol{\mu}_{\mathbf{c}_{t}}, \mathbf{\Sigma}_{\mathbf{c}_{t}}), \boldsymbol{\mu}_{\mathbf{c}_{t}}$ and $\mathbf{\Sigma}_{\mathbf{c}_{t}}$. The decoder takes the concatenation of the sampled $\mathbf{z}$ and the sampled $\mathbf{c}_{t}$ from their approximated posteriors as the input, and generates the parameters of $p_{\theta}(x \mid z)$ as the output.

\subsection{Loss function -- {\em ELBO+}}
\label{sec:elboplus}
With the variational attention mechanism, the variational lower bound $\mathcal{L}\left(\theta, \phi ; \mathbf{x}\right)$ in Eq.~\ref{eqn: elbo} becomes:

\begin{equation}
\begin{aligned}
\label{eqn:elbo_vrae}
\mathcal{L}(\theta, \phi, \boldsymbol{x})= E_{\boldsymbol{z}, \boldsymbol{c} \sim q_{\phi}\left(\boldsymbol{z}, \boldsymbol{c} \mid \boldsymbol{x}\right)}\left[\log p_{\theta}\left(\boldsymbol{x} \mid \boldsymbol{z}, \boldsymbol{c}\right)\right]\\
-\mathcal{D}_{\mathrm{KL}}\left(q_{\phi}\left(\boldsymbol{z}, \boldsymbol{c} \mid \boldsymbol{x}\right) \| p(\boldsymbol{z}, \boldsymbol{c})\right) 
\end{aligned}
\end{equation}
To minimise the effects of learning from abnormal data, we mitigate the contribution of global anomalies (pre-detected) and missing points by introducing a weighted vector, $\boldsymbol{\beta} = \left\{\beta_{i} \right\}_{i=1}^{T_{x}}$, to $\log p_{\theta}\left(\boldsymbol{x} \mid \boldsymbol{z}, \boldsymbol{c}\right)$  shown in Eq.~\ref{eqn:m_elbo_vrae}. If $x_i$ is an anomaly, then $\beta_i=0$, otherwise $\beta_i=1$. We name this improved ELBO as ELBO+, where $\lambda_{kl}$ weights the reconstruction loss and the KL loss and $\eta_{a}$ weights the latent KL loss and the attention KL loss. The training objective is to maximise the ELBO in Eq.~\ref{eqn:m_elbo_vrae}, which is the negative of the loss function for VAE. Theoretically, the anomalies present can also influence the KL losses, but the hyperparameters $\lambda_{kl}$ and $\eta_{a}$ can reduce the ratio of KL losses. We therefore do not reduce their contribution to the KL loss. 

\begin{equation}
\vspace{-10pt}
\begin{aligned}
\label{eqn:m_elbo_vrae}
\mathcal{L}(\theta, \phi, \boldsymbol{x})^+= E_{\boldsymbol{z} \sim q_{\phi}^{(z)}\left(\boldsymbol{z} \mid \boldsymbol{x}\right), \boldsymbol{c_t} \sim q_{\phi}^{(a)}\left(\boldsymbol{c_t} \mid \boldsymbol{x}\right)}\left[\boldsymbol{\beta} \log p_{\theta}\left(\boldsymbol{x} \mid \boldsymbol{z}, \boldsymbol{c}\right)\right] \\
-\lambda_{kl}\left[{ \mathcal{D}_{\mathrm{KL}}\left(q_{\phi}^{(z)}\left(\boldsymbol{z} \mid \boldsymbol{x}\right) \| p(\boldsymbol{z})\right)} \right. \\
\left. {-\eta_{a} \sum_{t=1}^{T} \mathcal{D}_{\mathrm{KL}}\left(q_{\phi}^{(a)}\left(\boldsymbol{c_t} \mid \boldsymbol{x}\right) \| p(\boldsymbol{c_t})\right) }\right] \\
\end{aligned}
\end{equation}

\subsection{Anomaly detection}
Since the generative model reconstructs the input time series based on the probability distribution, it can derive different outputs according to the probability distribution. Normally, rare events (anomalies) have lower probabilities. The rarity of events can be measured by the reconstruction probability, $ \log p_{\theta}\left(\boldsymbol{x} \mid \boldsymbol{z}\right)$, which can be calculated  through the Monte Carlo method. 

The encoder first generates the parameters of the approximate posterior distribution $\log p_{\phi}\left(\boldsymbol{z} \mid \boldsymbol{x}\right)$ using the test data. Then, sampled latent variables ($L$ samples) are derived from the approximate posterior distribution. The sampling strategy for latent variables takes into account the variability of the latent space in order to increase the robustness of anomaly detection. For each sample, the decoder outputs the parameters of the approximate posterior distribution $\log p_{\theta}\left(\boldsymbol{x} \mid \boldsymbol{z}\right)$. In the end, the average reconstruction probability of each sample is calculated from the output parameters, i.e.,
\begin{equation}
\begin{aligned}
\label{eqn:rp}
E_{\boldsymbol{z} \sim q_{\phi}\left(\boldsymbol{z} \mid \boldsymbol{x}\right))} \left[\log p_{\theta}\left(\boldsymbol{x} \mid \boldsymbol{z}\right)\right] \approx \frac{1}{L} \sum_{l=1}^{L} \log p_{\theta}\left(\boldsymbol{x} \mid \boldsymbol{\mu}_l, \boldsymbol{\sigma}_l \right)
\end{aligned}
\end{equation}
where $ \boldsymbol{\mu}_l$ and $\boldsymbol{\sigma}_l$ are the parameters as the output from the decoder for the approximate distribution $\log p_{\theta}\left(\boldsymbol{x} \mid \boldsymbol{z}\right)$.

The reconstruction probabilities are then used as anomaly scores (between 0 and 1), which measure the strength of the anomaly of input values. We consider the observations whose anomaly score is greater than 0.5 as the contextual anomalies for the experiments in the next section. This value can be tuned as per the requirement of the problem.

\section{Experiments}
\subsection{Data sets and experimental setup}
We use two data sets for the experiments, a synthetic data set generated by PyOD for the detection performance evaluation and a real smart meter data set about water supply temperature for district heating. In the real smart meter data set, we segment the time series into subsequences by a sliding window with a length of 168, and divide the subsequences into a training set, a validation set and a test set with a ratio of 75/15/10. We use PyTorch v1.6.0 to implement our algorithm and train the models via CPU i9-9900 and NVIDIA GTX 2080 Ti graphics cards with 16G RAM on Ubuntu 16.04. More details are included in appendices.

\subsection{Evaluation by comparison with baselines}
To evaluate our method, we compare it with 4 traditional methods and VAE-baseline using the synthetic data set, and use Precision ({\em P}), Recall ({\em R}) and F1 score ({\em F1}) as the metrics for the comparison. The comparing methods includes Cluster-based Local Outlier Factor (CBLOF) \cite{he2003discovering}, K nearest Neighbors (KNN) \cite{angiulli2002fast}, Principal component analysis (PCA) \cite{shyu2003novel}, One-class support vector machines (OCSVM) \cite{manevitz2001one} and VAE-baseline \cite{kingma2013auto}. The generated time series data by PyOD have 24,000 data points with 5 features, including 20\% abnormal data points. The normal 20,000 points are used for training and 4,000 points for testing.Table \ref{table:parameters} shows the hyperparameters used in our model.

From the results in Table~\ref{table:per1}, we can see that our method outperforms all others and are more effective for anomaly detection in terms of all three metrics. In general, the VAE-based networks have shown better performance in learning normal patterns from the train set. It also confirms that recurrent neural networks have a good capability to model long temporal dependency of time series.

\begin{table}[htp]
\vspace{-5pt}
\begin{minipage}[t]{0.5\textwidth}
\centering
\caption{Model hyperparameters}
\vspace{-5pt}
\begin{tabular}{p{4cm}p{1cm}}
\toprule  
LSTM hidden layers & 2 \\
Units in hidden layers & 218 \\
Sequence length $\left(W\right)$ & 168 \\
Latent dimensions & 3 \\
Training iterations & 550 \\
Learning rate & 0.0001 \\
Batch size & 1024 \\
Optimiser & Adam \\
$\lambda_{kl}$ & 0.01 \\
$\eta_a$ & 0.01 \\
\bottomrule
\end{tabular}
\label{table:parameters}
\end{minipage}
\begin{minipage}[t]{0.5\textwidth}
\centering
\caption{Performance on the synthetic dataset}
\vspace{-5pt}
\begin{tabular}{p{2cm}p{1cm}p{1cm}p{1cm}}
\toprule Method & \multicolumn{3}{c} { Metrics } \\
\cline { 2 - 4 } & $P$ & $R$ & $F1$ \\
\midrule
CBLOF & 0.65 & 0.68 & 0.66 \\
KNN & 0.69 & 0.69 & 0.69 \\
PCA & 0.83 & 0.84 & 0.83 \\
OCSVM &0.82 & 0.82&0.82 \\
VAE-baseline & 0.89 & 0.90 & 0.90 \\
\textbf{Ours} & \textbf{0.95} & \textbf{0.93} & \textbf{0.94} \\
\bottomrule
\end{tabular}
\label{table:per1}
\end{minipage}
\vspace{-15pt}
\end{table}

\subsection{An empirical case study}
We next evaluate the proposed method by an empirical case study, which detects the ``anomalies" in the water temperature time series from an industry district heating company. The data is recorded from 19/09/2019 11:05:00 to 11/08/2020 15:00:00 with 23 sensors in irregular minute-level, with a total of 220,097 observations for each time series. We align these fine-grained readings to hourly resolution by aggregation, and obtain 7,582 observations for each time series.

\vspace{-20pt}
\begin{figure}
  \begin{minipage}[t]{0.5\linewidth}
    \centering
    \includegraphics[scale=0.5]{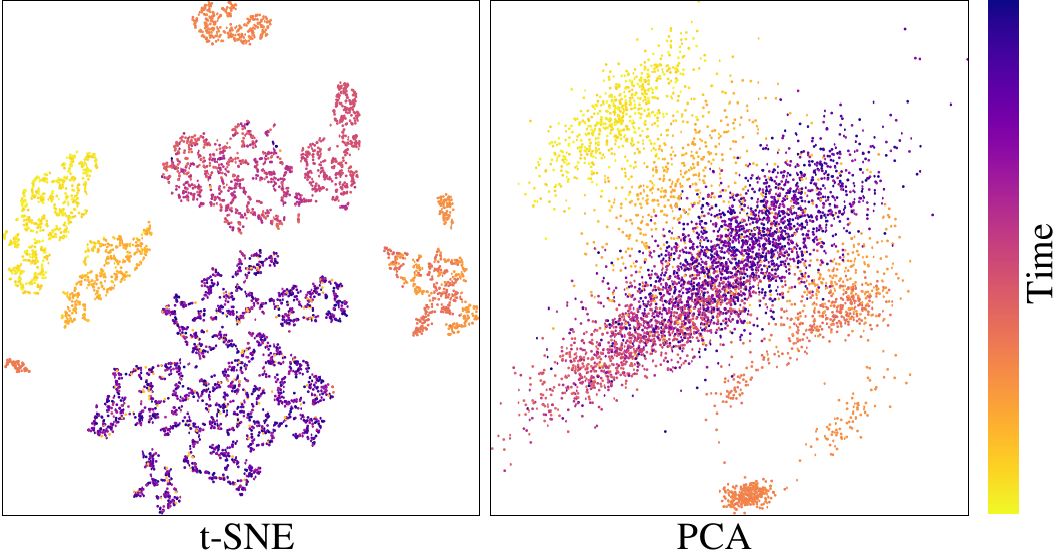}
    \vspace{-12pt}
    \caption{Latent space visualisation of training set}
    \label{fig:z}
  \end{minipage}%
  \begin{minipage}[t]{0.5\linewidth}
    \centering
    \includegraphics[scale=0.30]{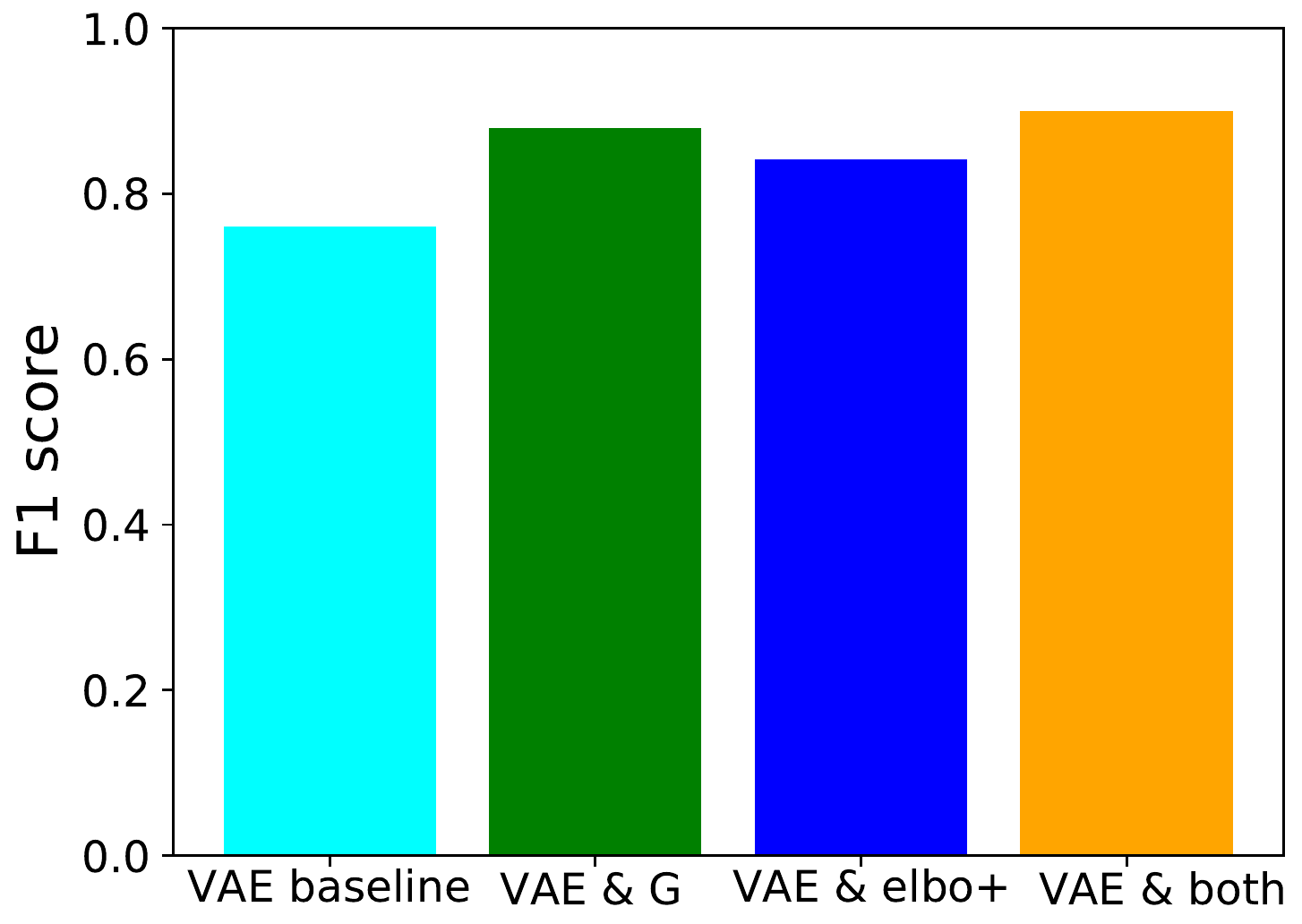}
    \vspace{-12pt}
    \caption{F1 score of different models}
    \label{fig:elbo1}
  \end{minipage}
\end{figure}
\vspace{-20pt}

For visualisation purposes, we reduce the 3-dimensional latent variables of the training set to 2D and visualise them using Principal Component Analysis (PCA) and t-distributed Stochastic Neighbour Embedding (t-SNE) \cite{maaten2008visualizing}. Figure~\ref{fig:z}  shows the projected points of subsequences in the latent space by the dimensional reduction methods, t-SNE and PCA, respectively. The more similar subsequence are, the closer these points are placed. The color legend represents the time from the beginning (bottom) to the end (top) sequences.

\subsubsection{Effects of ELBO+ and pre-detected global anomalies}
To our knowledge, minimising the impact of anomalies during training can assist the learning processing of networks. To exam the effectiveness of the pre-detected global anomalies and our modified ELBO loss function, we calculate F1 score in the test set of the real smart meter data set to compare the performance under different conditions. The F1 score is a measure of test accuracy and calculated from the precision and recall of the test. The four models are (1) VAE baseline, (2) VAE with global anomaly detection, (3) VAE with elbo+, and (4) VAE with both global anomaly detection and elbo+. Figure \ref{fig:elbo1} shows that predetected global anomalies and elbo+ have a positive effect on anomaly detection and our model outperforms the VAE baseline. 

\begin{figure}
  \begin{minipage}[t]{0.5\linewidth}
    \centering
    \includegraphics[scale=0.42]{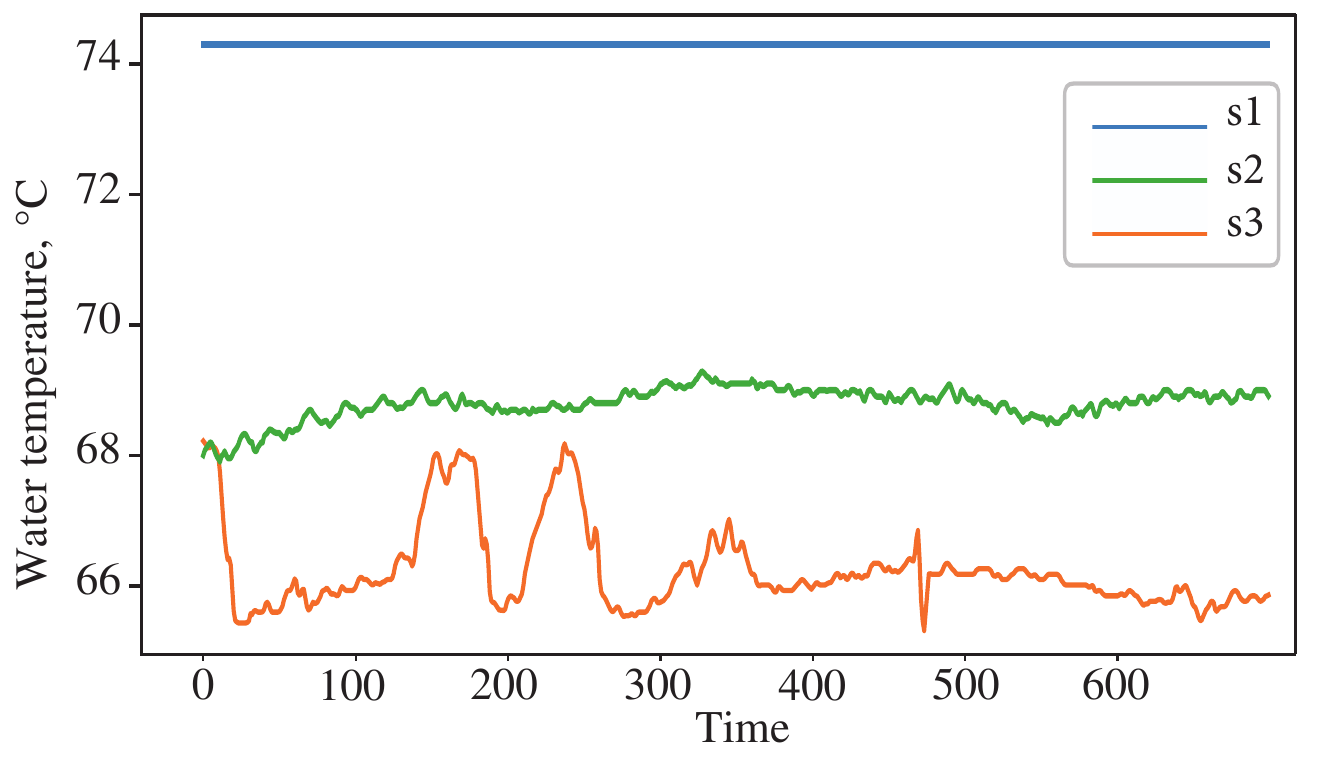}
    \vspace{-14pt}
    \caption{Distinctive features (or patterns) in hot water temperature time series.}
    \label{fig:features}
  \end{minipage}%
  \hspace{.02\linewidth}
  \begin{minipage}[t]{0.5\linewidth}
    \centering
    \includegraphics[scale=0.3]{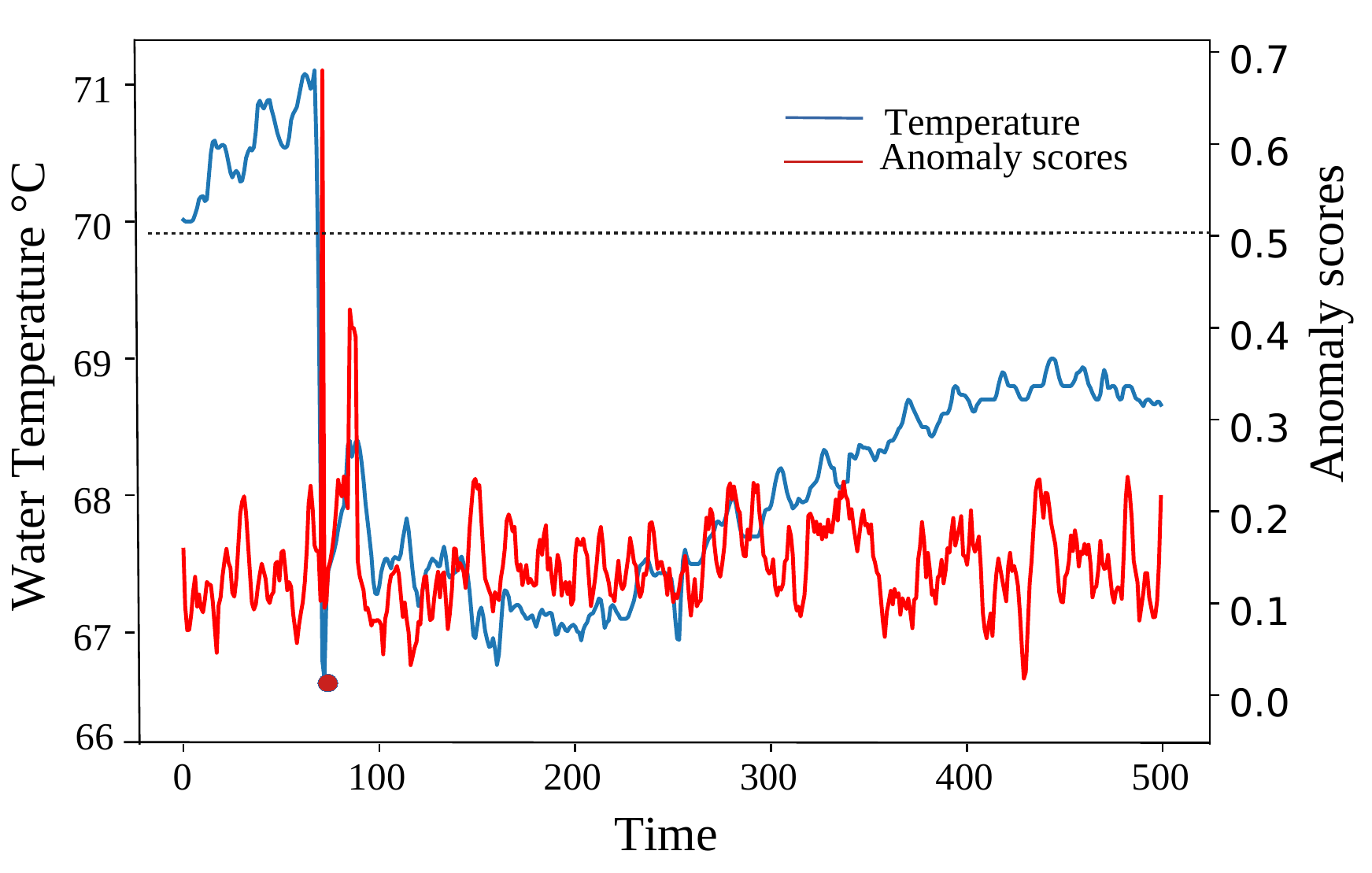}
    \vspace{-14pt}
    \caption{An example of a contextual anomaly with corresponding anomaly scores.}
    \label{fig:local}
  \end{minipage}
\end{figure}

From Figure~\ref{fig:z}, the latent space has a clear tendency to group which implies there are distinct features between subsequences. Here, we gives three examples (Figure~\ref{fig:features}) of subsequences in a time series with distinctive features. From 19/09/2019 to 12/12/2019, the transmission water temperature for district heating continuously is 74.2\textdegree{}C, which is a straight line ($s1$). $s2$ is about 68.4\textdegree{}C and is stationary. By contrast, $s3$ has a lower temperature and is nonstationary. 

\vspace{-20pt}
\vspace{-20pt}

The model outputs reconstruction probabilities of time series as anomaly scores for each point. When the anomaly score is higher than 0.5, the corresponding point is classified as an anomaly. We give an example of detected contextual anomalies in hot water temperature dataset. Figure \ref{fig:local} shows a contextual anomaly example and the corresponding anomaly scores of points. The red dot indicates a contextual anomalies.

\section{Conclusions and future work}
In this paper, we proposed an unsupervised anomaly detection algorithm for smart meter data using VRAE with attention mechanism. Our method can detect different types of anomalies including global and contextual anomalies. The enhanced ELBO+ function can mitigate the contribution of global anomalies and missing points.  We have evaluated our method comprehensively and the results have demonstrated the effectiveness and superiority of our method. For future work, we would further improve our case study by applying a real-time architecture for online anomaly detection. We would also address dealing with concept drifts during the anomaly detection process.

\vspace{8pt}
\noindent {\bf Acknowledgements.} The research was supported by  Heat4.0 project (8090-00046A) and the project FlexSUS: Flexibility for Smart Urban Energy Systems
(91352) funded by the European Union's Horizon 2020 research and innovation programme under grant agreement No 77597. 
\bibliographystyle{splncs04}
{\small
\bibliography{ref1}

\begin{thebibliography}{10}
\providecommand{\url}[1]{\texttt{#1}}
\providecommand{\urlprefix}{URL }
\providecommand{\doi}[1]{https://doi.org/#1}

\bibitem{an2015variational}
An, J., Cho, S.: Variational autoencoder based anomaly detection using
  reconstruction probability. Special Lecture on IE  \textbf{2}(1) (2015)

\bibitem{angiulli2002fast}
Angiulli, F., Pizzuti, C.: Fast outlier detection in high dimensional spaces.
  In: European conference on principles of data mining and knowledge discovery.
  pp. 15--27. Springer (2002)

\bibitem{bahdanau2014neural}
Bahdanau, D., Cho, K., Bengio, Y.: Neural machine translation by jointly
  learning to align and translate. arXiv preprint arXiv:1409.0473  (2014)

\bibitem{bahuleyan2017variational}
Bahuleyan, H., Mou, L., Vechtomova, O., Poupart, P.: Variational attention for
  sequence-to-sequence models. arXiv preprint arXiv:1712.08207  (2017)

\bibitem{breunig2000lof}
Breunig, M.M., Kriegel, H.P., Ng, R.T., Sander, J.: Lof: identifying
  density-based local outliers. In: Proceedings of the 2000 ACM SIGMOD
  international conference on Management of data. pp. 93--104 (2000)

\bibitem{chahla2019novel}
Chahla, C., Snoussi, H., Merghem, L., Esseghir, M.: A novel approach for
  anomaly detection in power consumption data. In: ICPRAM. pp. 483--490 (2019)

\bibitem{2009Anomaly}
Chandola, V., Banerjee, A., Kumar, V.: Anomaly detection: A survey. Acm
  Computing Surveys  \textbf{41}(3),  1--58 (2009)

\bibitem{chen2011energy}
Chen, C., Cook, D.J.: Energy outlier detection in smart environments. In:
  Workshops at the Twenty-Fifth AAAI Conference on Artificial Intelligence
  (2011)

\bibitem{creswell2018generative}
Creswell, A., White, T., Dumoulin, V., Arulkumaran, K., Sengupta, B., Bharath,
  A.A.: Generative adversarial networks: An overview. IEEE Signal Processing
  Magazine  \textbf{35}(1),  53--65 (2018)

\bibitem{7836693}
{Deng}, J.D.: Online outlier detection of energy data streams using incremental
  and kernel pca algorithms. In: 2016 IEEE 16th International Conference on
  Data Mining Workshops (ICDMW). pp. 390--397 (2016)

\bibitem{fan2018analytical}
Fan, C., Xiao, F., Zhao, Y., Wang, J.: Analytical investigation of
  autoencoder-based methods for unsupervised anomaly detection in building
  energy data. Applied energy  \textbf{211},  1123--1135 (2018)

\bibitem{goodfellow2016deep}
Goodfellow, I., Bengio, Y., Courville, A., Bengio, Y.: Deep learning, vol.~1.
  MIT press Cambridge (2016)

\bibitem{he2003discovering}
He, Z., Xu, X., Deng, S.: Discovering cluster-based local outliers. Pattern
  Recognition Letters  \textbf{24}(9-10),  1641--1650 (2003)

\bibitem{hochreiter1997long}
Hochreiter, S., Schmidhuber, J.: Long short-term memory. Neural computation
  \textbf{9}(8),  1735--1780 (1997)

\bibitem{hollingsworth2018energy}
Hollingsworth, K., Rouse, K., Cho, J., Harris, A., Sartipi, M., Sozer, S.,
  Enevoldson, B.: Energy anomaly detection with forecasting and deep learning.
  In: 2018 IEEE International Conference on Big Data (Big Data). pp.
  4921--4925. IEEE (2018)

\bibitem{jakkula2010outlier}
Jakkula, V., Cook, D.: Outlier detection in smart environment structured power
  datasets. In: 2010 Sixth International Conference on Intelligent
  Environments. pp. 29--33. IEEE (2010)

\bibitem{kingma2013auto}
Kingma, D.P., Welling, M.: Auto-encoding variational bayes. arXiv preprint
  arXiv:1312.6114  (2013)

\bibitem{kriegel2009loop}
Kriegel, H.P., Kr{\"o}ger, P., Schubert, E., Zimek, A.: Loop: local outlier
  probabilities. In: Proceedings of the 18th ACM conference on Information and
  knowledge management. pp. 1649--1652 (2009)

\bibitem{5175339}
{Li}, X., {Bowers}, C.P., {Schnier}, T.: Classification of energy consumption
  in buildings with outlier detection. IEEE Transactions on Industrial
  Electronics  \textbf{57}(11),  3639--3644 (2010)

\bibitem{liu2016online}
Liu, X., Iftikhar, N., Nielsen, P.S., Heller, A.: Online anomaly energy
  consumption detection using lambda architecture. In: International Conference
  on Big Data Analytics and Knowledge Discovery. pp. 193--209. Springer (2016)

\bibitem{liu2020}
Liu, X., Lai, Z., Wang, X., Huang, L., Nielsen, P.S.: A contextual anomaly
  detection framework for energy smart meter data stream. In: International
  Conference on Neural Information Processing. pp. 733--742. Springer (2020)

\bibitem{liu2018}
Liu, X., Nielsen, P.S.: Scalable prediction-based online anomaly detection for
  smart meter data. Information Systems  \textbf{77},  34--47 (2018)

\bibitem{lund2017smart}
Lund, H., {\O}stergaard, P.A., Connolly, D., Mathiesen, B.V.: Smart energy and
  smart energy systems. Energy  \textbf{137},  556--565 (2017)

\bibitem{luong2015effective}
Luong, M.T., Pham, H., Manning, C.D.: Effective approaches to attention-based
  neural machine translation. arXiv preprint arXiv:1508.04025  (2015)

\bibitem{maaten2008visualizing}
Maaten, L.v.d., Hinton, G.: Visualizing data using t-sne. Journal of machine
  learning research  \textbf{9}(Nov),  2579--2605 (2008)

\bibitem{malhotra2016lstm}
Malhotra, P., Ramakrishnan, A., Anand, G., Vig, L., Agarwal, P., Shroff, G.:
  Lstm-based encoder-decoder for multi-sensor anomaly detection. arXiv preprint
  arXiv:1607.00148  (2016)

\bibitem{malhotra2015long}
Malhotra, P., Vig, L., Shroff, G., Agarwal, P.: Long short term memory networks
  for anomaly detection in time series. In: Proceedings. vol.~89, pp. 89--94.
  Presses universitaires de Louvain (2015)

\bibitem{manevitz2001one}
Manevitz, L.M., Yousef, M.: One-class svms for document classification. Journal
  of machine Learning research  \textbf{2}(Dec),  139--154 (2001)

\bibitem{pereira2018unsupervised}
Pereira, J., Silveira, M.: Unsupervised anomaly detection in energy time series
  data using variational recurrent autoencoders with attention. In: 2018 17th
  IEEE International Conference on Machine Learning and Applications (ICMLA).
  pp. 1275--1282. IEEE (2018)

\bibitem{pol2019anomaly}
Pol, A.A., Berger, V., Germain, C., Cerminara, G., Pierini, M.: Anomaly
  detection with conditional variational autoencoders. In: 2019 18th IEEE
  International Conference On Machine Learning And Applications (ICMLA). pp.
  1651--1657. IEEE (2019)

\bibitem{santolamazza2018anomaly}
Santolamazza, A., Cesarotti, V., Introna, V.: Anomaly detection in energy
  consumption for condition-based maintenance of compressed air generation
  systems: An approach based on artificial neural networks. IFAC-PapersOnLine
  \textbf{51}(11),  1131--1136 (2018)

\bibitem{scholkopf1998nonlinear}
Sch{\"o}lkopf, B., Smola, A., M{\"u}ller, K.R.: Nonlinear component analysis as
  a kernel eigenvalue problem. Neural computation  \textbf{10}(5),  1299--1319
  (1998)

\bibitem{seem2007using}
Seem, J.E.: Using intelligent data analysis to detect abnormal energy
  consumption in buildings. Energy and buildings  \textbf{39}(1),  52--58
  (2007)

\bibitem{shyu2003novel}
Shyu, M.L., Chen, S.C., Sarinnapakorn, K., Chang, L.: A novel anomaly detection
  scheme based on principal component classifier. Tech. rep., MIAMI UNIV CORAL
  GABLES FL DEPT OF ELECTRICAL AND COMPUTER ENGINEERING (2003)

\bibitem{solch2016variational}
S{\"o}lch, M., Bayer, J., Ludersdorfer, M., van~der Smagt, P.: Variational
  inference for on-line anomaly detection in high-dimensional time series.
  arXiv preprint arXiv:1602.07109  (2016)

\bibitem{su2019robust}
Su, Y., Zhao, Y., Niu, C., Liu, R., Sun, W., Pei, D.: Robust anomaly detection
  for multivariate time series through stochastic recurrent neural network. In:
  Proceedings of the 25th ACM SIGKDD International Conference on Knowledge
  Discovery \& Data Mining. pp. 2828--2837 (2019)

\bibitem{vaswani2017attention}
Vaswani, A., Shazeer, N., Parmar, N., Uszkoreit, J., Jones, L., Gomez, A.N.,
  Kaiser, {\L}., Polosukhin, I.: Attention is all you need. In: Advances in
  neural information processing systems. pp. 5998--6008 (2017)

\bibitem{xu2018unsupervised}
Xu, H., Chen, W., Zhao, N., Li, Z., Bu, J., Li, Z., Liu, Y., Zhao, Y., Pei, D.,
  Feng, Y., et~al.: Unsupervised anomaly detection via variational auto-encoder
  for seasonal kpis in web applications. In: Proceedings of the 2018 World Wide
  Web Conference. pp. 187--196 (2018)

\bibitem{zhang2019time}
Zhang, C., Chen, Y.: Time series anomaly detection with variational
  autoencoders. arXiv preprint arXiv:1907.01702  (2019)

\bibitem{zhang2019deep}
Zhang, C., Song, D., Chen, Y., Feng, X., Lumezanu, C., Cheng, W., Ni, J., Zong,
  B., Chen, H., Chawla, N.V.: A deep neural network for unsupervised anomaly
  detection and diagnosis in multivariate time series data. In: Proceedings of
  the AAAI Conference on Artificial Intelligence. vol.~33, pp. 1409--1416
  (2019)

\bibitem{zhang2011anomaly}
Zhang, Y., Chen, W., Black, J.: Anomaly detection in premise energy consumption
  data. In: 2011 IEEE Power and Energy Society General Meeting. pp.~1--8. IEEE
  (2011)

\bibitem{zhao2014adaptive}
Zhao, J., Liu, K., Wang, W., Liu, Y.: Adaptive fuzzy clustering based anomaly
  data detection in energy system of steel industry. Information Sciences
  \textbf{259},  335--345 (2014)

\bibitem{6520712}
{Zhao}, Y., {Lehman}, B., {Ball}, R., {Mosesian}, J., {de Palma}, J.: Outlier
  detection rules for fault detection in solar photovoltaic arrays. In: 2013
  Twenty-Eighth Annual IEEE Applied Power Electronics Conference and Exposition
  (APEC). pp. 2913--2920 (2013)

\bibitem{zhao2019pyod}
Zhao, Y., Nasrullah, Z., Li, Z.: Pyod: A python toolbox for scalable outlier
  detection. arXiv preprint arXiv:1901.01588  (2019)

\end{thebibliography}
}

\appendix
\section{Appendix}
\subsection{Data sets}

We use two types of data sets for the experiments: a synthetic data set and a real smart meter data set. We use the PyOD toolkit \cite{zhao2019pyod} to generate synthetic data sets with anomalies. The normal data were sampled from a multivariate Gaussian distribution, while the anomalies were sampled from a uniform distribution. The generated time series data have 24,000 data points with five features, including 20\% abnormal data points. The normal 20,000 points are used for training and 4,000 points for testing. 

The smart meter data are the supply hot water temperature provided by a district heating company in Denmark. As there are no labels in the time series, the anomalies are detected by the unsupervised learning method. The data used are from 19/09/2019 11:05:00 to 11/08/2020 15:00:00. The time series data before 04/12/2019 09:00 have an hourly resolution, but have an irregular minute-level resolution after that time, ranging from 1 to 5 minutes, with a total of 220,097 observations. Therefore, we align these fine-grained readings to hourly resolution by aggregation, and obtain a total of 7,582 observations. 

\subsection{Experimental setup}

\begin{table}[htp!]
\vspace{-8pt}
\centering
\begin{tabular}{p{4.5cm}p{3cm}}
\toprule 
LSTM hidden layers & 2 \\
Units in hidden layers & 218 \\
Sequence length $\left(W\right)$ & 168 \\
Latent dimensions & 3 \\
Training iterations & 550 \\
Learning rate & 0.0001 \\
Batch size & 1024 \\
Optimiser & Adam \\
Input dimensions & 5 (synthetic data), 23 (real data) \\
Divergence ratio $\lambda_{kl}$ & 0.01 \\
Attention divergence ratio $\eta_a$ & 0.01 \\
\bottomrule 
\end{tabular}
\caption{Model hyperparameters}
\label{table:parameters}
\vspace{-10pt}
\end{table}

Table~\ref{table:parameters} shows the model parameters used in the experiments. We use the window size $W=168$, as a smaller size may not be able to capture the normal patterns, while a larger size may increase the risk of overfitting. We set the latent space to 3 dimensions for visualization purposes. The encoder and decoder consist of two hidden LSTM layers, 218 units in each layer. We have tested several combinations of hyperparameters, including the number of the hidden LSTM layers, the number of hidden units in each layer, and the sequence length. However, the results do not show much difference in terms of the loss. The number of training iterations is determined based on the convergence of training loss and validation loss.

\end{document}